\documentclass[10pt]{article}
\usepackage[OE]{express}

\usepackage{graphicx}
\usepackage[ruled,vlined]{algorithm2e}
\usepackage{amsmath,amssymb,amsfonts}       
\DeclareMathOperator{\sinc}{sinc}

\newcommand{\FTd}{\mathop{\mathbf{TF}}}
\newcommand{\iFTd}{\mathop{\mathbf{TF}^{-1}}}

\newcommand{\FTR}{\widetilde{\mathcal{R}}} 
\newcommand{\FTG}{\widetilde{\mathcal{G}}} 
\newcommand{\FTS}{{\tilde{\mathbf{s}}}} 

\newcommand{\svec}{{\mathbf{s}}}

\newcommand{\etavec}{{\boldsymbol{\eta}}} 

\SetCommentSty{mycommfont}

\begin{document}

\title{Iterative wave-front reconstruction in the Fourier domain}

\author{Charlotte Z. Bond,\authormark{1*} Carlos M. Correia,\authormark{1} Jean-Fran\c{c}ois Sauvage,\authormark{1,2}
Benoit Neichel,\authormark{1} and Thierry Fusco\authormark{1,2}}

\address{\authormark{1}Aix Marseille Universite, CNRS, LAM (Laboratoire d'Astrophysique de Marseille) UMR 7326, 13388, Marseille, France \\
\authormark{2}ONERA (Office National d'Etudes et de Recherches A\'erospatiales), B.P.72, F-92322 Ch\^atillon, France}

\email{\authormark{*}charlotte.bond@lam.fr}

\begin{abstract}
The use of Fourier methods in wave-front reconstruction can significantly reduce the computation time
for large telescopes with a high number of degrees of freedom.  
However, Fourier algorithms for 
discrete data require a rectangular data set which conform to specific boundary requirements,
whereas wave-front sensor data is typically defined over a circular domain (the telescope
pupil).
Here we present an iterative Gerchberg routine modified for the purposes of
discrete wave-front reconstruction which adapts the measurement data (wave-front sensor slopes)
for Fourier analysis, fulfilling the requirements of the Fast Fourier Transform (FFT) 
and providing accurate reconstruction.
The routine is used in the adaptation step
only and can be coupled to any other Wiener-like or least-squares method.
We compare simulations using this method with previous Fourier
methods and show an increase in performance in terms of Strehl ratio
and a reduction in noise propagation for a 40$\times$40 SPHERE-like adaptive optics system.
For closed loop operation with minimal iterations
the Gerchberg method provides an improvement in Strehl, 
from 95.4\% to 96.9\% in K-band.  This corresponds to $\sim40$\,nm
improvement in rms, and avoids the high spatial frequency errors present in other methods,
providing an increase in contrast towards the edge of the correctable band.
\end{abstract}


\section{Introduction}

The performance of large ground-based telescopes depends on
the ability of adaptive optics (AO) systems to correct for phase distortions
caused by atmospheric turbulence.  This is highly dependent on the
accuracy of the wave-front measurements provided by the AO systems
wave-front sensor and the speed at which the wave-front is measured
and corrected.  Therefore an accurate high speed wave-front reconstruction 
algorithm is highly desirable.

Future Extremely Large Telescopes (ELTs) are in the design process~\cite{ELT14,TMT07,GMT12}.
These large facilities will incorporate complex AO systems with a very
high number of degrees of freedom, specifically the number of
correctable spatial modes $\sim 5 \times 10^3$.
To provide a high rate of correction (in the kHz regime) very fast wave-front reconstruction
algorithms are required.  Standard wavefront reconstruction requires $N\times N$
operations, with $N$ being the number of correctable modes to solve
for the linear system of equations that is customarily used.
Use of Fast Fourier Transforms (FFTs) can in principle reduce the number of operations
to $N\log(N)$, providing a vast improvement in the speed of the reconstruction
process~\cite{Montilla10}.  For an ELT sized system this is a potential reduction
in the number of operations by 3 orders of magnitude.

%

Initial attempts to use Fourier methods for wave-front 
reconstruction~\cite{Frost79, Freischlad86b} did not take into account
the telescope pupil imposed on the data, applying such methods
to data defined over a rectangular domain.  In~\cite{Roddier91} the issue of an arbitrary data
boundary, such as the pupil, is addressed with the proposition of a Gerchberg-like
algorithm to extrapolate the known data beyond the boundary and
simultaneously reconstruct the phase. 
The routine employs the use of the Laplacian of the data, $\nabla^2
= \partial^2/\partial x^2 + \partial^2/\partial y^2$, in the
continuous Fourier
domain leading to 
a simple, complex multiplication for reconstruction of the phase
$\tilde{\hat{\phi}} = \left(\kappa_x\FTS_x + \kappa_y
  \FTS_y\right)/(2i\pi|\kappa|^2)$ where $\tilde{\hat{\phi}}$ is the
  reconstructed phase in Fourier space, $\kappa$ is the
  spatial-frequency vector and $\FTS$ are slopes in Fourier space.
However, this fails to take into account a more realistic measurement
process, particularly the discrete nature of the wave-front
measurement in terms of lenslet sampling and the pixelated
WFS spots on the detector, and to justify the use of the Laplacian as the solution of
a least-squares minimisation of the reconstruction process~\cite{Correia14}.
Additional works have developed other Fourier methods for wave-front
reconstruction based on satisfying the boundary conditions of
the FFT~\cite{Poyneer02,Poyneer03,Ribak06}. 

In this paper we expand on the methods presented in~\cite{Roddier91, Correia08},
reconstructing the wave-front using Fourier filters derived from discrete
models of the measurement process, as developed in~\cite{Correia14}.  We include additional
constraints, specifically enforcing periodicity on the WFS data
and demonstrate the improvements offered by this routine 
using open and closed loop end-to-end simulations of an adaptive
optics system using a Shack-Hartmann WF sensor.
The specific motivation is to provide a robust reconstruction
in the Fourier domain in order to investigate anti-aliasing filters developed in~\cite{Correia14}.
To observe the anticipated gains in contrast we require accurate reconstructors
which perform well across the correction band, and particularly avoid
errors at high spatial frequencies.

\section{Reconstruction in the Fourier domain}
\label{Sec:FFTR}
The description of the wave-front measurement process can be approximated
by a series of convolutions in direct space (the \emph{forward model}).
In the Fourier domain this translates to a series of multiplications \cite{Correia14}:
\begin{equation}
\FTS(\kappa) = \tilde{G}\tilde{\phi}(\kappa) + \tilde{\etavec}
\end{equation}
where $\kappa = \{\kappa_x,\kappa_y\}$ is the bi-dimensional spatial frequency vector, $\FTS$ is the slope
measurement, $\tilde{G}$ describes the slope measurement
for given phase, $\tilde{\phi}$ is the phase
and $\tilde{\eta}$ is the measurement noise.  $\tilde{}$ denotes a
variable in the Fourier space.  

Wave-front reconstruction in the Fourier domain is achieved
using filters, $\tilde{R}$, which convert the measured slopes into 
a reconstructed phase:
\begin{equation}
\tilde{\hat{\phi}} = \tilde{R}_x \FTS_x + \tilde{R}_y \FTS_y
\end{equation}
The inverse Fourier transform of $\tilde{\hat{\phi}}$ gives the
reconstructed phase in direct space, $\hat{\phi}$.
The construction of $\tilde{R}$ depends on the measurement model used and
various filters have previously been constructed based on different
wave-front sensor geometries~\cite{Poyneer05,Freischlad86,Correia14,Correia08}.  
The general form of such a filter is a Wiener filter, the Fourier equivalent
of a direct space minimum-mean-squared-error (MMSE) method,
which minimises the residual error variance and maximises the Strehl:
\begin{equation}\label{eq:WienerFilter}
\tilde{R} = \frac{\tilde{G}^{*}}{|\tilde{G}|^2 + \gamma\frac{W_{\eta}}{W_{\phi}}}
\end{equation}
$W_{\eta}$ is the power spectral density (PSD) of the noise
in the measurement and $W_{\phi}$ is the PSD of the turbulent
phase.  * represents complex conjugate variables.
$\gamma$ is a constant with $\gamma=0$ representing a least squares (LSQ) reconstruction
and $\gamma=1$ the MMSE.
In the results presented here we consider the MMSE filter.

Equation~(\ref{eq:WienerFilter}) shows that the solution for Fourier reconstruction involves
the division by $|\tilde{G}|^2$, which in continuous space is equal to
$\kappa_x^2 + \kappa_y^2$, and equivalent to the formulation involving the Laplacian 
proposed by Roddier\&Roddier \cite{Roddier91}.  We consider the filters
developed in~\cite{Correia14} which take into account the discrete measurement
process.  These correspond to the following direct space measurement model
of the Shack-Hartmann:
\begin{equation}
\label{eq:G}
G = {\bf |||}\left(\frac{{\bf x}}{d}\right) \times \left[\Pi\left(\frac{{\bf x}}{d}\right) 
\otimes \nabla  \right]
\end{equation}
where $d$ is the lenslet spacing, $\nabla$ is the gradient function,
$\times$ represents a simple multiplication and
$\otimes$ is a convolution.  The comb function, ${\bf |||}\left(\frac{{\bf x}}{d}\right)$,
represents the lenslet sampling and manifests itself as an aliasing factor in
the Fourier domain.  The convolution with the square function, $\Pi\left(\frac{{\bf x}}{d}\right)$,
represents the process by which the gradients are averaged in
the Shack-Hartmann measurement: each measurement is the difference between
the average phase across the opposite edges of the lenslet.  A
subtle correction to the formalism of Eq.~(\ref{eq:G}) as derived in~\cite{Correia14}
is that the difference is effectively taken over $d-d_{px}$, where $d_{px}$ is the
length of one pixel on the WFS detector.  This averaging process is represented
in the Fourier domain as multiplications by $\sinc$ functions.

\section{Adapting to the aperture of the telescope}

\subsection{Limitations of Fourier analysis for non-rectangular domains}

The Fourier transform decomposes a function into Fourier modes, with 
these modes satisfying the conditions of
an orthonormal basis over an infinite cartesian space.
Analysis of discrete data defined over a finite space, such as slope
measurements from a WFS, requires the Fast Fourier Transform (FFT).
The Fourier modes used in the FFT are defined by the interval and sampling
of the data.  The FFT requires data defined over a rectangular space and
assumes that the data is periodic.  In this way the data can be completely
described by the FFT modes.  If either of these two conditions are not full-filled
errors can occur in the Fourier analysis.

In the case of adaptive optics systems the data is the slope measurements
from a wave-front sensor.  The telescope aperture is superimposed on this
data and therefore valid measurements do not fill a rectangular space.  
The sharp cut-off imposed by the aperture is interpreted by numerical FT 
algorithms as additional frequency components which do not represent the frequency
content of the slope measurement but rather the convolution, in Fourier space,
with the pupil function.  This leads to large errors in the frequency content
which propagate through the reconstruction process,
resulting in large errors in the estimated phase~\cite{Poyneer02}.

\subsection{Extending wave-front sensor data}
\label{Sec:ext}
To avoid the errors induced by the imposition of the pupil the wave-front
sensor data must be extended outside the aperture, conforming
to the periodic conditions required by
the FFT and the characteristics of the wave-front sensor data~\cite{Freischlad86b,Freischlad93}.
These conditions can be characterised by two constraints:
\begin{enumerate}
\item $\frac{1}{D} \int_D \nabla \phi \ \mathrm{d}{\bf x}  = 0 $.
If the phase is spatially periodic then the integral of the gradients across the data set must
be zero.  Numerically this means the sum of each row (for $x$ slopes) and 
each column (for $y$ slopes) is equal to zero.
This can be imposed during the slope extension as follows~\cite{Poyneer02}:
\begin{equation}
\label{eq:PC}
   \begin{array}{cc}
      s_x(m,N) = -\sum_{n=1}^{N-1}s_x(m,n) &   1 \leq m \leq N \\
       s_y(N,n) = -\sum_{m=1}^{N-1}s_y(m,n) &   1 \leq n \leq N \\
   \end{array}
\end{equation}

\item $rot(\nabla \phi) = 0$.
If the phase is spatial continuous then the curl of the gradient is zero and any closed
path sum of gradients must equal zero.
\end{enumerate}
As both constraints are derived from a continuous noise-less approach they may be impacted by 
noise in the measurement and the discrete nature of the measurement.

Two methods proposed in~\cite{Poyneer02} suggest extending the data
by extrapolating the gradients at the edges of the aperture, conforming
to the constraints identified above.
The \emph{extension method} defined in~\cite{Poyneer02}, here referred to
as the \emph{Hudgin extension}, extends the slopes by repeating the gradients
at the edge of the aperture until the rectangular data boundary is reached.  
The $x$ slopes are extended up and down from
the aperture and the $y$ slopes are extended left and right.
The periodic condition is then explicitly imposed.  In this paper
periodicity is always enforced when
the Hudgin extension is used.

Methods of extending the data such as the Hudgin extension have
demonstrated good reconstruction and an avoidance of the large errors
due to the imposition of the aperture.  However, they only extrapolate from
the data at the edges which will be most impacted by noise on account
of partial illumination of the sub-apertures. Our aim is to devise an extension method which 
extrapolates the overall characteristics of the data within the aperture.
Here we propose a new extension method based on an iterative Gerchberg
process in the Fourier domain, as previously suggested in~\cite{Correia08}.

\subsection{Gerchberg method}
\label{sec:gerchMeth}
The Gerchberg routine is designed to extend known data outside the
measurement bounds, using Fourier analysis to extrapolate the
data and retain the Fourier properties of the known data over an extended region.
Originally proposed by Gerchberg in 1974~\cite{Gerchberg74}
as a method for extending spectra to enhance resolution, it can be applied
to WFS data to extend beyond the bounds of the aperture.
An instructive illustration of the Gerchberg method applied to
1D data is shown in~\cite{Chatterjee07}.

For our purposes the Gerchberg routine has been adapted for wave-front reconstruction
using the Fourier measurement filters.
The filter replaces the band-width restriction in the general Gerchberg algorithm~\cite{Chatterjee07}
(the low pass filter).
The first step takes the Fourier transform of the raw slope data. 
Then the phase is reconstructed in Fourier space:
\begin{equation}
\hat{\tilde{\phi}} = \frac{\tilde{G}^*_x\FTS_x + \tilde{G}^*_y\FTS_y}{|\tilde{G}_x|^2 + |\tilde{G}_y|^2}
\end{equation}
In this step the phase is reconstructed using a filter equivalent to a least squares approach
(i.e. no noise PSD component in the denominator, see Eq.~(\ref{eq:WienerFilter})).  
Within the Gerchberg routine the filter is applied for each iteration.
Inclusion of the noise term (equivalent to an MMSE approach) for recursive filtering
leads to an over-filtering of the noise and potential large errors in the extended data.
Subsequently the MMSE filter is only used for the final phase reconstruction.

Using the measurement filter ($\tilde{G}$) the slopes are estimated from the 
reconstructed phase and an inverse Fourier transform returns the estimated slopes
in direct space:
\begin{equation}
\hat{\tilde{s}} = \tilde{G} \hat{\tilde{\phi}} \ \ \ \ \longrightarrow \ \ \ \ \hat{s} = \mathfrak{F}^{-1}(\hat{\tilde{s}})
\end{equation}
The final slopes estimate takes the original measurement within the pupil
and the estimated slopes outside the pupil.
The Gerchberg process is then repeated as necessary, with the possibility of
enforcing the periodic condition on the final slopes.  A simplified algorithm is
as follows:

\begin{algorithm}[H]
  \KwIn{WFS measurements: $\{\svec_x^0; \svec_y^0\}$}
  \For{ $i = 1,2,...,\,n_{\mathrm{iter.}} $  }{
    1: Fourier transform $\rightarrow$ slopes in Fourier space \\
    $\FTS_{x}^{i-1} = \FTd \{\svec_x ^{i-1}\}$\\
    $\FTS_{y}^{i-1} = \FTd \{\svec_y^{i-1}\}$ \\
    2: Phase reconstruction (filtering) followed by gradient computation \\
    $\FTS_{x}^i  = \FTG_x  \left(\FTR_x \ \FTS_{x}^{i-1} + \FTR_y \ \FTS_{y}^{i-1}\right)$\\
    $\FTS_{y}^i= \FTG_y \left(\FTR_x \ \FTS_{x}^{i-1} + \FTR_y \ \FTS_{y}^{i-1}\right)$

    3: Inverse Fourier transform $\rightarrow$ extended slopes in direct space\\
    $\svec_x^i = \iFTd\{\FTS_{x}^i\} $\\
    $\svec_y^i = \iFTd\{\FTS_{y}^i\} $\\
    4: Replace gradients within the pupil by the original measurement, keeping extrapolated gradients
    	outside the pupil\\ 
    $\svec_x^i=  \svec_x^0 P + \svec_x^i \overline{P}$\\
    $\svec_y^i =  \svec_y^0 P + \svec_y^i\overline{P}$
   }
   5: If required the periodic condition is then enforced\\
   \If{$P_{\mathrm{cond.}} == \mathrm{true}$}{
      $\svec_x(m,N) = -\sum_{n=1}^{N-1}\svec_x(m,n) $ \\
       $\svec_y(N,n) = -\sum_{m=1}^{N-1}\svec_y(m,n) $ \\
   }
  \KwOut{Gradients over a $N \times N$
    computational grid}
  \caption{Modified Gerchberg algorithm in which slope measurements
  	are extended beyond the aperture.  The gradients
    obtained are periodic over the domain and the rotational
    is as close to zero as possible on account of the
    measurement noise.}
  \label{alg:myGerchberg}
\end{algorithm}
%

Whilst with enough iterations ($n_{\mathrm{iter.}}$) the Gerchberg routine will 
result in extended slope data which satisfies
the periodic nature of the FFT the Hudgin extension requires explicit enforcement
of the periodic condition (using the approach described in Eq.~(\ref{eq:PC})),
as this method of extending only the values at the aperture edge does not
satisfy periodicity.
The slopes initially generated through the Gerchberg routine are the result of an FFT
and will therefore satisfy the conditions of the FFT (step 3 of the algorithm).  The replacement of
the data within the pupil by the original measurement results in a new set of non-periodic
slopes (step 4), but with enough iterations the data will conform to
a periodic data set.  If a large number of iterations are required the periodic
condition can be explicitly enforced as with the Hudgin extension.

\begin{figure}[htp]
    \centerline{
      \includegraphics[scale=0.25]{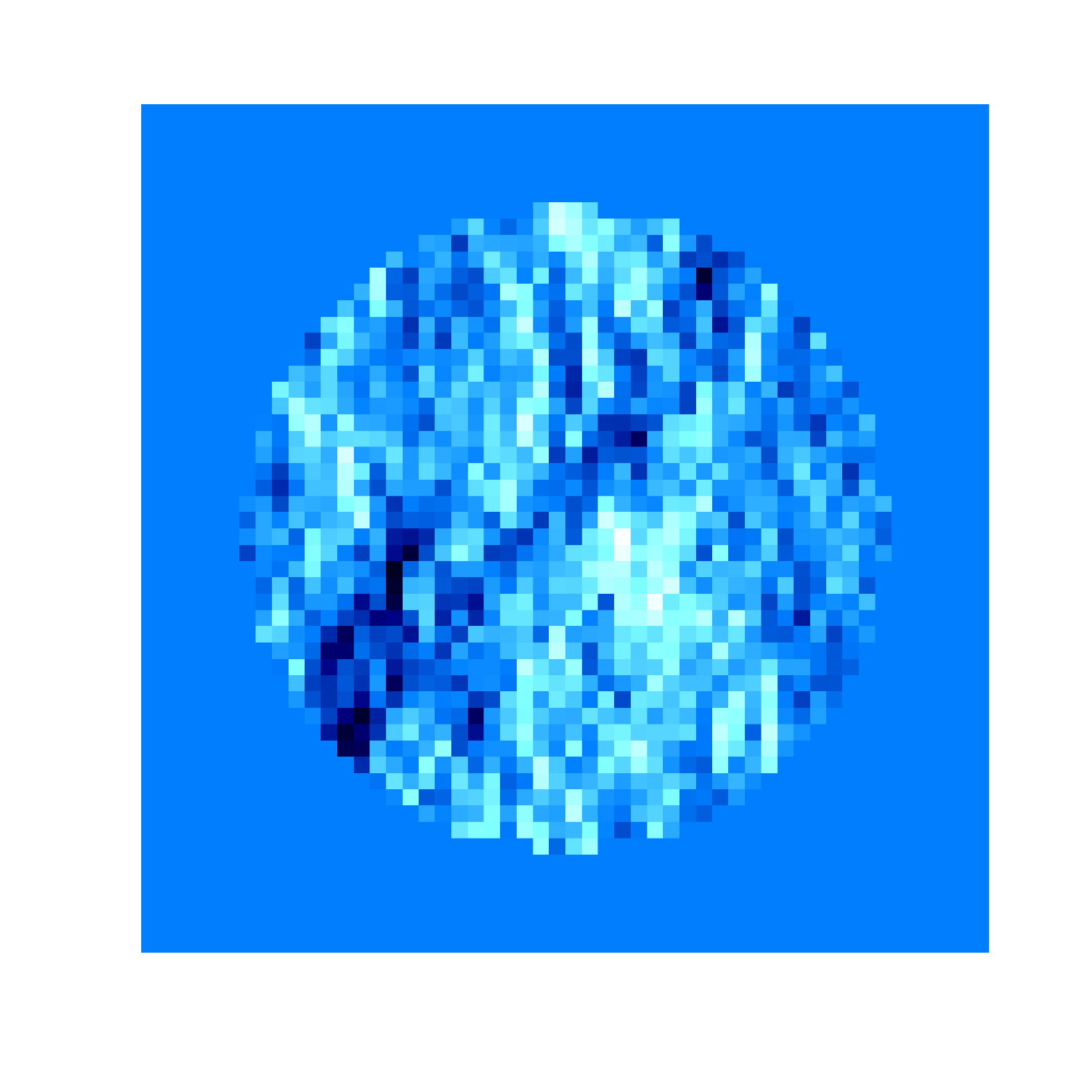}
      \includegraphics[scale=0.25]{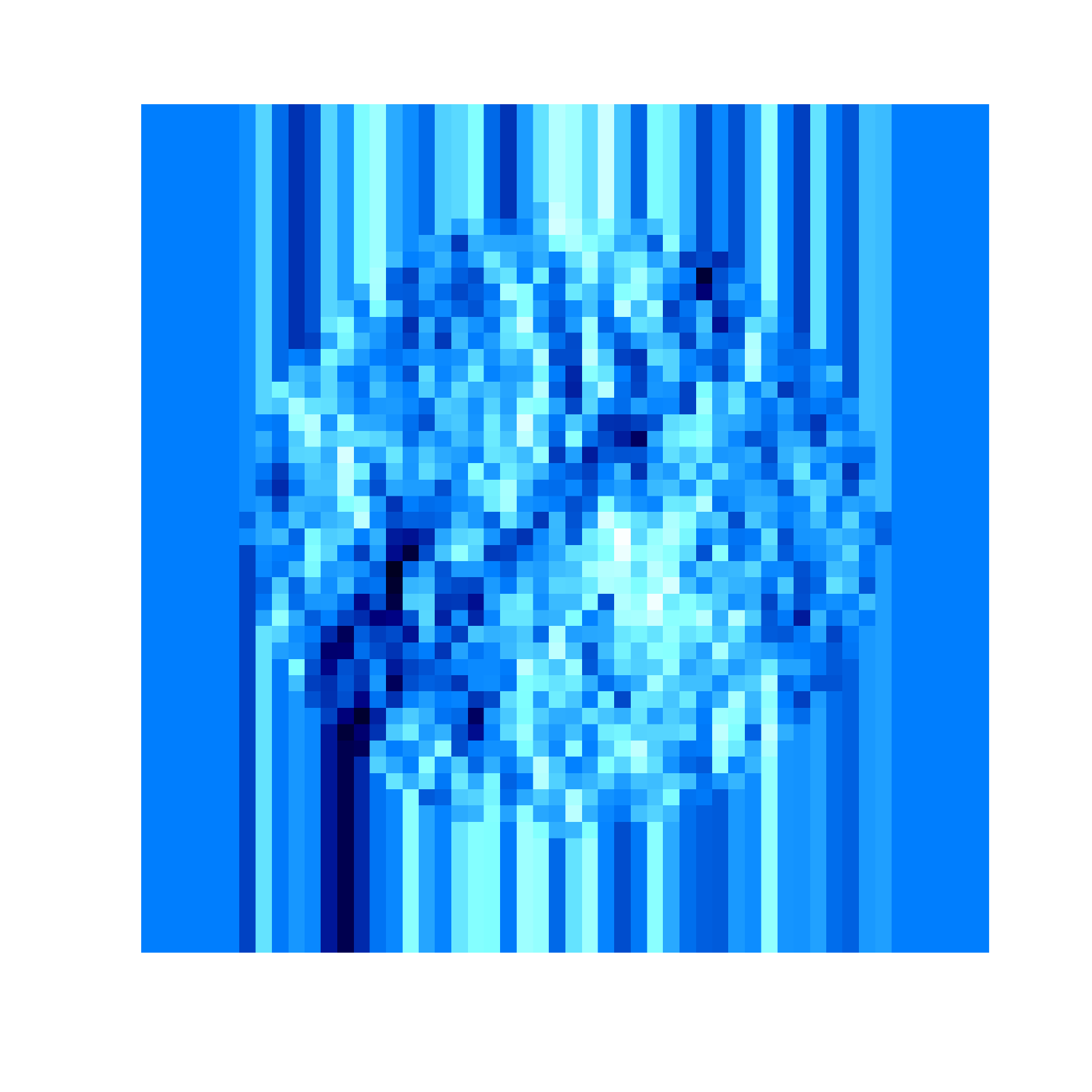}
      \includegraphics[scale=0.25]{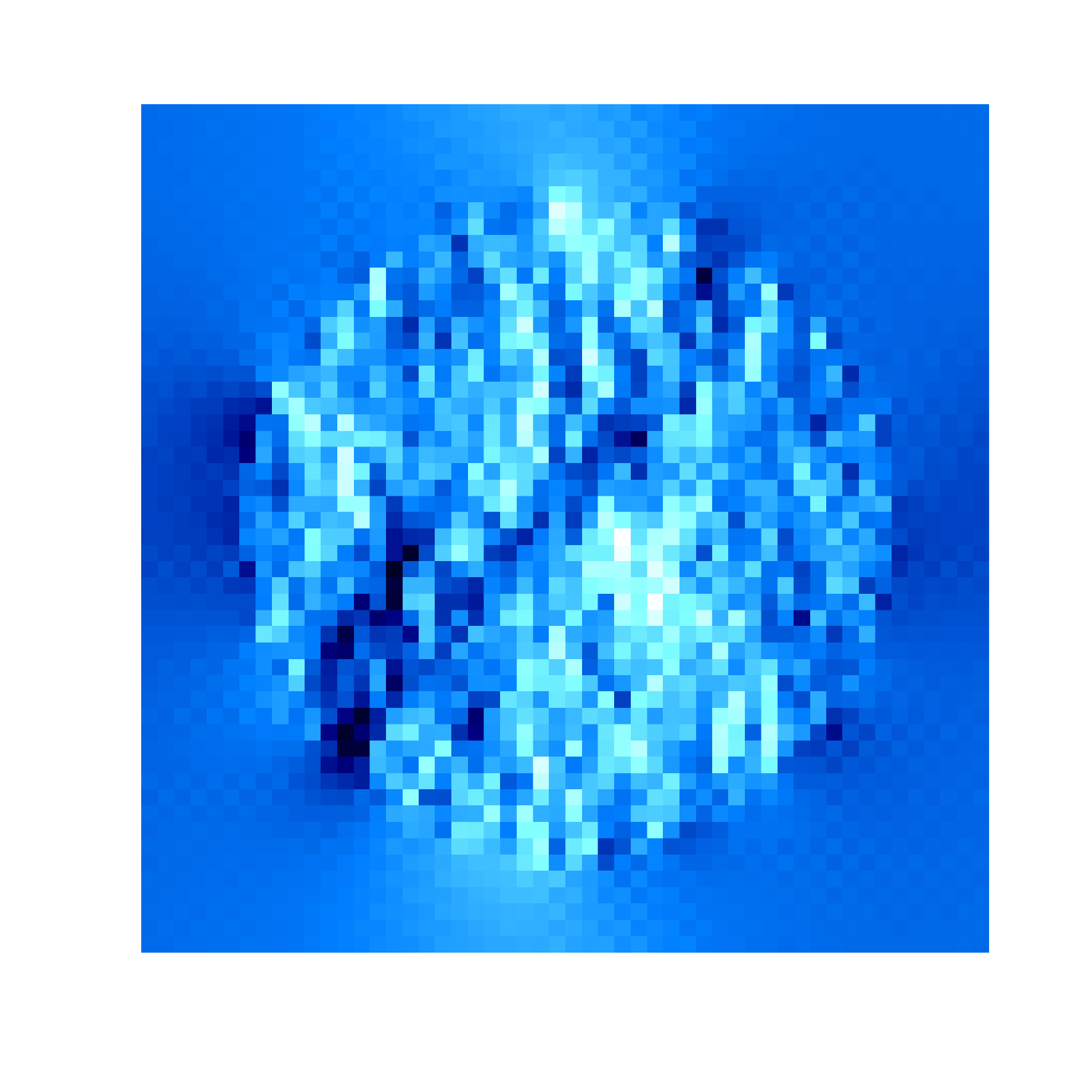}
    }
    \caption{Examples of different extension methods.  {\bf Left:} Original $x$ slopes.
    		{\bf Middle:} $x$ slopes extended using Hudgin extension method.
		{\bf Right:} Extended $x$ slopes using the Gerchberg method with 3 iterations.
		The examples are shown here before/without the imposition of the periodic condition.}
    \label{fig:extendedSlopes}
\end{figure}

Figure~\ref{fig:extendedSlopes} shows an example of the Gerchberg extension
(right) compared to the Hudgin extension (middle) and the original slope
measurement (left). The Gerchberg extension exhibits a more gradual change
in data across the boundary imposed by the aperture, whilst both the original
slopes and Hudgin extended slopes contain sharper features.

\begin{figure}[htp]
    \centerline{
      \includegraphics[scale=0.6]{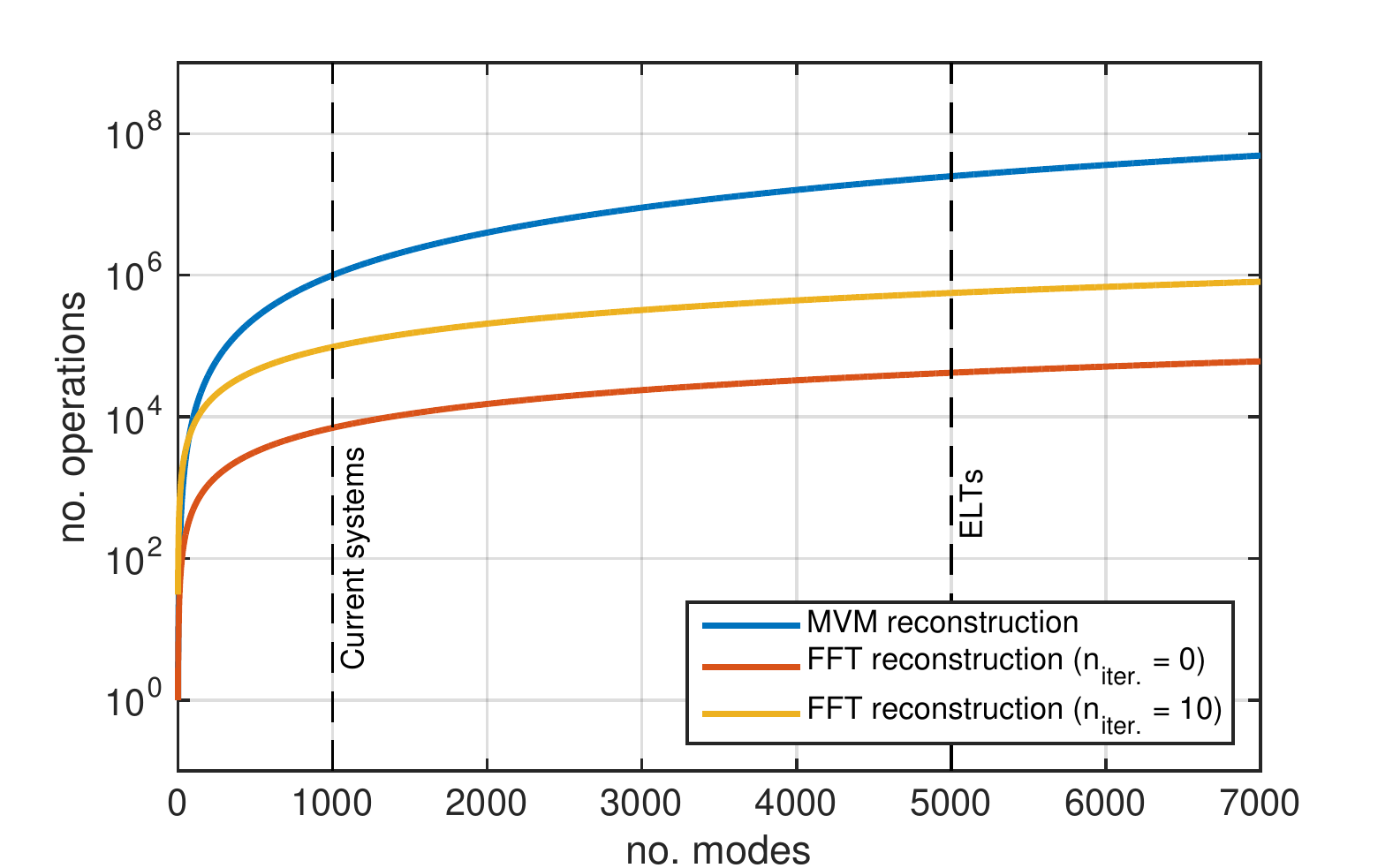}
    }
    \caption{Plot illustrating the number of operations required for wave-front
    	reconstruction of an $N$ mode system, both in direct space ($N\times N$) 
	and for possible FFT reconstruction (see Eq.~(\ref{eq:Nop})).  For FFT
	reconstruction two cases are shown, one without the Gerchberg extension 
	($n_{\mathrm{iter.}}=0$) and one with 10 Gerchberg iterations ($n_{\mathrm{iter.}}=10$).
	The state for current systems and 
	ELTs are highlighted, showing the potential for a 1 -- 2 order of magnitude reduction
	in the number of operations using Fourier methods.}
    \label{fig:noOpps}
\end{figure}

The use of the Gerchberg routine will result in an increase in the number of 
operations.  For each set of slopes the number of operations for the reconstruction process
are as follows:
\begin{equation}
\label{eq:Nop}
N_{\mathrm{op.}} = 
\underbrace{n_{\mathrm{iter.}} (2N\log(N) + 3N)}_{\text{Gerchberg algorithm}} + 
\underbrace{2N\log(N)+N}_{\text{Final reconstruction}}
\end{equation}
where $n_{\mathrm{iter.}}$ is the number of Gerchberg iterations and $N$ is
the number of system modes (or number of grid points).  The $2N\log(N)$ terms
refer to the forwards and backwards FFTs (in both the Gerchberg algorithm and
final reconstruction).  In the Gerchberg algorithm the $3N$ operations refer to the
phase reconstruction, the gradient computation and the replacement of the
original measurement, whilst the final reconstruction term
only includes one $N$ term for the phase reconstruction.
This equation should be compared with the standard number of operations for
a direct space matrix vector multiply method (MVM), $N^2$.  In Fig.~\ref{fig:noOpps}
the number of operations for FFT reconstruction and MVM reconstruction are
compared, with and without application of the Gerchberg method.  Even with a large
number of iterations of the Gerchberg algorithm ($n_{\mathrm{iter.}}=10$)
the FFT reconstruction will offer a significant reduction in the number of operations
required for ELTs ($>$ 2 orders of magnitude with a small number of iterations).

We envisage the possible advantages of the Gerchberg method as producing
more accurate results in both open and closed loop.  However, if a large
number of iterations are required this will increase computing time,
diminishing the increase in speed offered by the FFT, whilst also
adding another layer of complexity to the reconstruction process,
providing greater possibility for noise propagation through the algorithm.

\section{End-to-end simulation results}

In order to test the performance of the Gerchberg extension
we compare the results of end-to-end simulations using our extension
method and the Hudgin extension.  
Particularly we aim to investigate the number of iterations required to achieve
a comparable performance, the impact of enforcing periodicity with the
Gerchberg extension and the effect of the extension window (number of
pixels either side of the aperture).

The results presented here are for a SPHERE-like case~\cite{Fusco06}
and the parameters used are summarised in table~\ref{tab:simParams}.
\begin{table}[h]
\caption{Summary of simulation parameters used in end-to-end simulations of
		a full AO loop.  The atmospheric properties (the Fried parameter $r_0$ and
		outer scale $L_0$) are both given, as well as the telescope properties and
		wavelengths for the wave-front sensing and science star.
		The parameters correspond to a SPHERE-like system with the
		omission of the spatial filter before the WFS~\cite{Fusco06}.}
\begin{center}
\begin{tabular}{l|l}
Parameter 				& Value	\\
\hline
$r_0$					& 15\,cm	\\
$L_0$					& 30\,m	\\
No. atmosphere layers		& 3		\\
Telescope diameter			& 8\,m	\\
No. lenslets				& 40		\\	
No. pixels per lenslet			& 6		\\		
$\lambda_{\mathrm{WFS}}$	& V band \\
$\lambda_{\mathrm{sci.}}$	& K band \\
Read-out noise				& 2\,$e^{-}$ 	\\
Guide star magnitude		& 8			\\
Median wind speed			& 10\,ms$^{-1}$ \\
Sampling frequency			& 1\,kHz	\\
\end{tabular}
\end{center}
\label{tab:simParams}
\end{table}%
The simulations are carried out using OOMAO (Object Oriented
Matlab Adaptive Optics toolbox) a \textsc{Matlab} based simulation
code~\cite{Conan14}.  The performance is quantified in terms of 
the spatial frequency content
of the residual phase, the Strehl and noise propagation (as defined in~\cite{Zou06}).
The performance is compared for both open and closed loop cases
and for simplicity a deformable mirror is not included (to separate
DM fitting errors from the impacts of the different extension methods).
The simulated atmosphere includes higher spatial frequency components than
those defined by the lenslet spacing and therefore, with the omission of
the spatial filter present in the SPHERE system, the simulations will include aliasing errors.

In all cases presented here the filter used for the final phase
reconstruction and within the Gerchberg algorithm is
the Rigaut filter, as detailed in~\cite{Correia14}.
The extension of the slopes data is carried out using three different methods:
\begin{enumerate}
\item The Gerchberg extension.
\item The Gerchberg extension with periodic condition enforced (using Eq.~(\ref{eq:PC})).
\item The Hudgin extension (always enforces periodic condition).
\end{enumerate}

\newpage
\subsection{Open loop}
\label{sec:OL}
Firstly we consider the open loop case.  When no extension method
is applied the errors caused by the aperture and non-periodic nature
of the slope measurement result in a low Strehl ratio of 25.3\,\% in K band.
This can be compared to a theoretical Strehl of $\sim97\%$,
estimated using the Marechal approximation and including fitting
and aliasing errors~\cite{Rigaut98}.

\begin{figure}[t]
    \centerline{
      \includegraphics[scale=0.55]{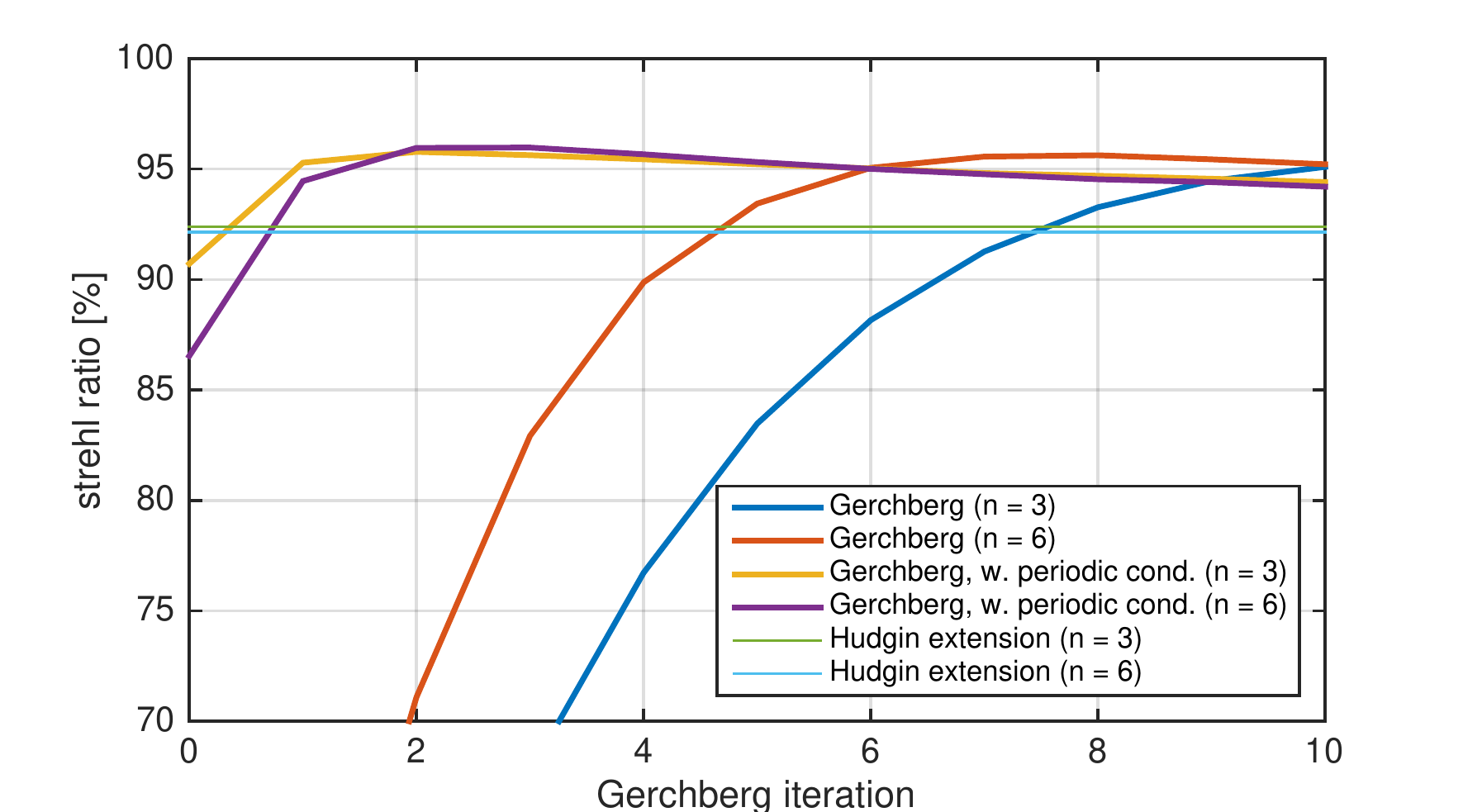}
    }
    \caption{Plots of the K-band Strehl ratio versus Gerchberg iteration for
    		open loop end-to-end simulations.  The results for
		2 different extension windows ($n=3$ and $n=6$) are shown
		for 3 extension methods: 1) Gerchberg extension with enforced periodicity;
		2) Gerchberg extension without enforced periodicity; and 3) the Hudgin
		extension.}
    \label{fig:OL_SRvsIter}
\end{figure}

Figure~\ref{fig:OL_SRvsIter} shows the Strehl ratio obtained using the Gerchberg
extension versus iteration.  The results are shown for different extension windows
($n=3$ and $n=6$ pixels) as well as with and without enforcement of the periodic condition 
(see Sec.~\ref{Sec:ext} and~\cite{Poyneer02}).
Results using the Hudgin extension are also shown.
With enough iterations all instances of the Gerchberg extension perform 
better than the Hudgin extension.  However, enforcing periodicity on the 
data is crucial to achieving a good
Strehl within a limited number (1 or 2) Gerchberg iterations.
Without this the system takes significantly more iterations to converge.

A peak in performance using the Gerchberg routine is observed
(i.e. at 2 iterations with the periodic condition).  Additional iterations
then result in a slight drop in Strehl.
This can be the consequence of trying to fit the measured data to
our model (the Fourier filters) through successive iterations,
where the filters are not an exact match to
the true measurement process.  Another factor is an increase
in noise propagation with additional iterations, as discussed in more detail below.

Figure~\ref{fig:OL_SRvsIter} also reveals a difference in performance depending
on the size of the extension window ($n=3$ and $n=6$).  We 
observe this even in the case of no slope extension
(i.e. 0 Gerchberg iterations corresponds to padding the original data with zeros).
When the periodic condition is enforced the Strehl is already significantly improved.
Without the periodic condition the Strehl is $\sim30\%$ ($n=3$) and $\sim35\%$ ($n=6$)
compared to 25.3\% for the non-extended case.  The impact of the extension window
is discussed in greater detail in Sec.~\ref{sec:CL}.

\begin{figure}[t]
    \centerline{
      \includegraphics[scale=0.55]{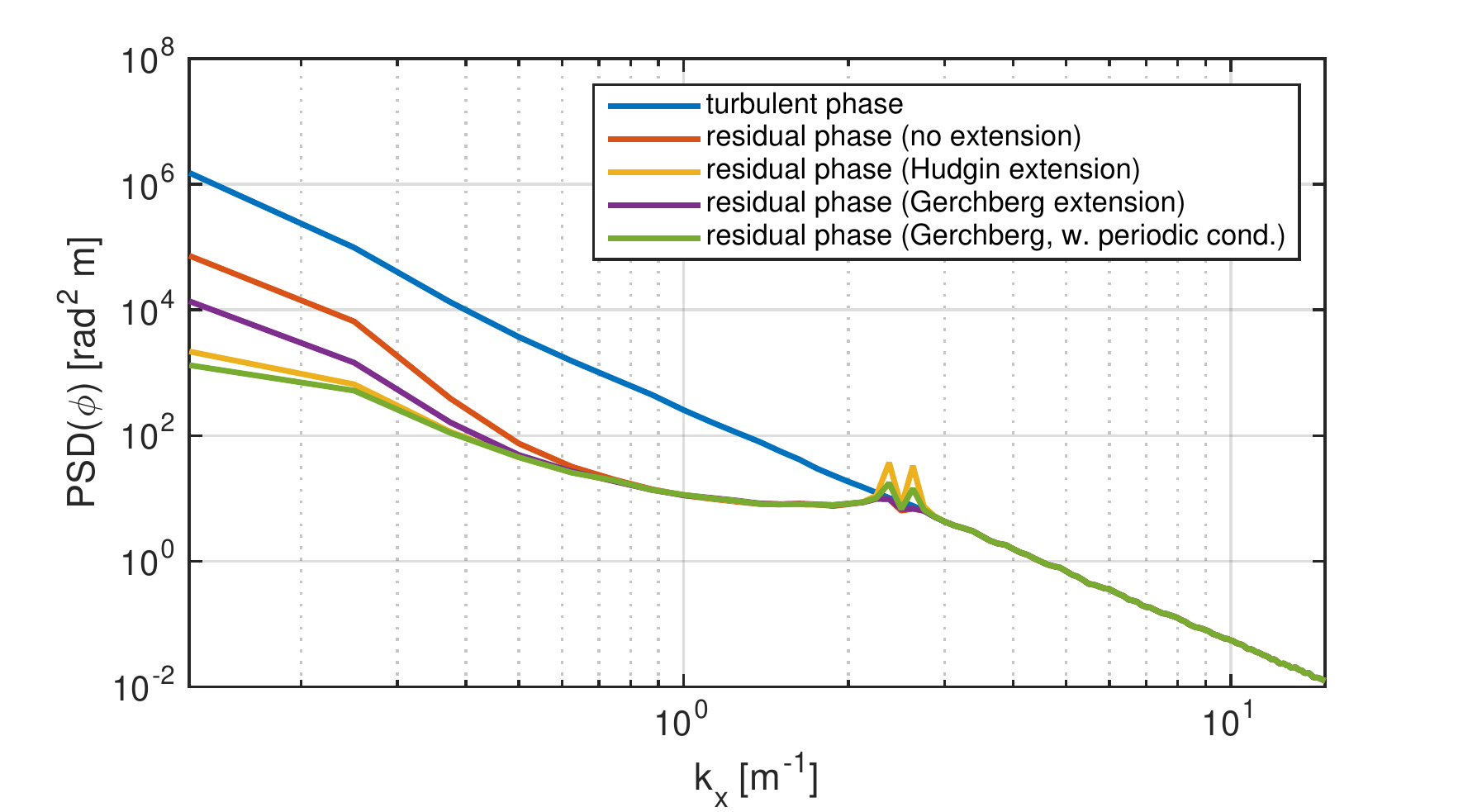}
    }
    \caption{PSDs (power spectral densities) of the turbulent phase and
    		residual phase for different open loop simulations.  In all cases
		the phase is reconstructed in the Fourier domain using different 
		extension methods: 1) no extension; 2) Hudgin extension; 3)
		Gerchberg extension (3 iterations); and 4) Gerchberg extension with
		periodicity enforced (3 iterations).  In the 3 extension cases (2\,--\,4)
		an extension window of $n=3$ is used.}
    \label{fig:OL_PSDs}
\end{figure}

Although enforcing periodicity results in a greater Strehl with
fewer iterations the consequence is an error at higher
frequencies, specifically near the waffle frequency
($\frac{1}{2d} = 2.5$\,m$^{-1}$).
This is illustrated in the power spectral densities (PSDs)
of the residual phase shown in Fig.~\ref{fig:OL_PSDs}.
For the case with no extension (red trace) the PSD shows
a relatively high low frequency content.  This is the result of the
cut-off imposed by the telescope aperture: large
errors in the estimation of the low spatial frequencies.
The 3 extensions shown here all improve on the case with no extension.
The two Gerchberg cases here use 3 iterations, in
order to compare performance for a sufficiently
low number of iterations.
In the two cases where periodicity is enforced (the Hudgin
extension and Gerchberg case) the low spatial frequency content
is significantly reduced compared to the no extension case,
with the Gerchberg extension doing slightly better than the Hudgin.
This is reflected in a significant improvement in Strehl.
However, at high spatial frequencies ($\sim 2.5$\,m$^{-1}$)
there is a small error in the reconstruction.  This is a result of
enforcing periodicity along a single row/column of the slopes data,
introducing a high spatial frequency element.  This is not present
in the Gerchberg case where periodicity is not enforced.
However, as discussed above, it is not desirable to use the large number of iterations
required for high Strehl when periodicity is not enforced
in the Gerchberg extension.
The cost of optimising the overall performance (the Strehl)
using the periodic condition is a small high spatial frequency error.
Such errors can then be filtered out of the reconstructed phase using a specific filter~\cite{Poyneer02,Poyneer03}.
In addition the DM will provide some additional correction
at these high frequencies~\cite{Correia14}.

\begin{figure}[htp]
    \centerline{
      \includegraphics[scale=0.55]{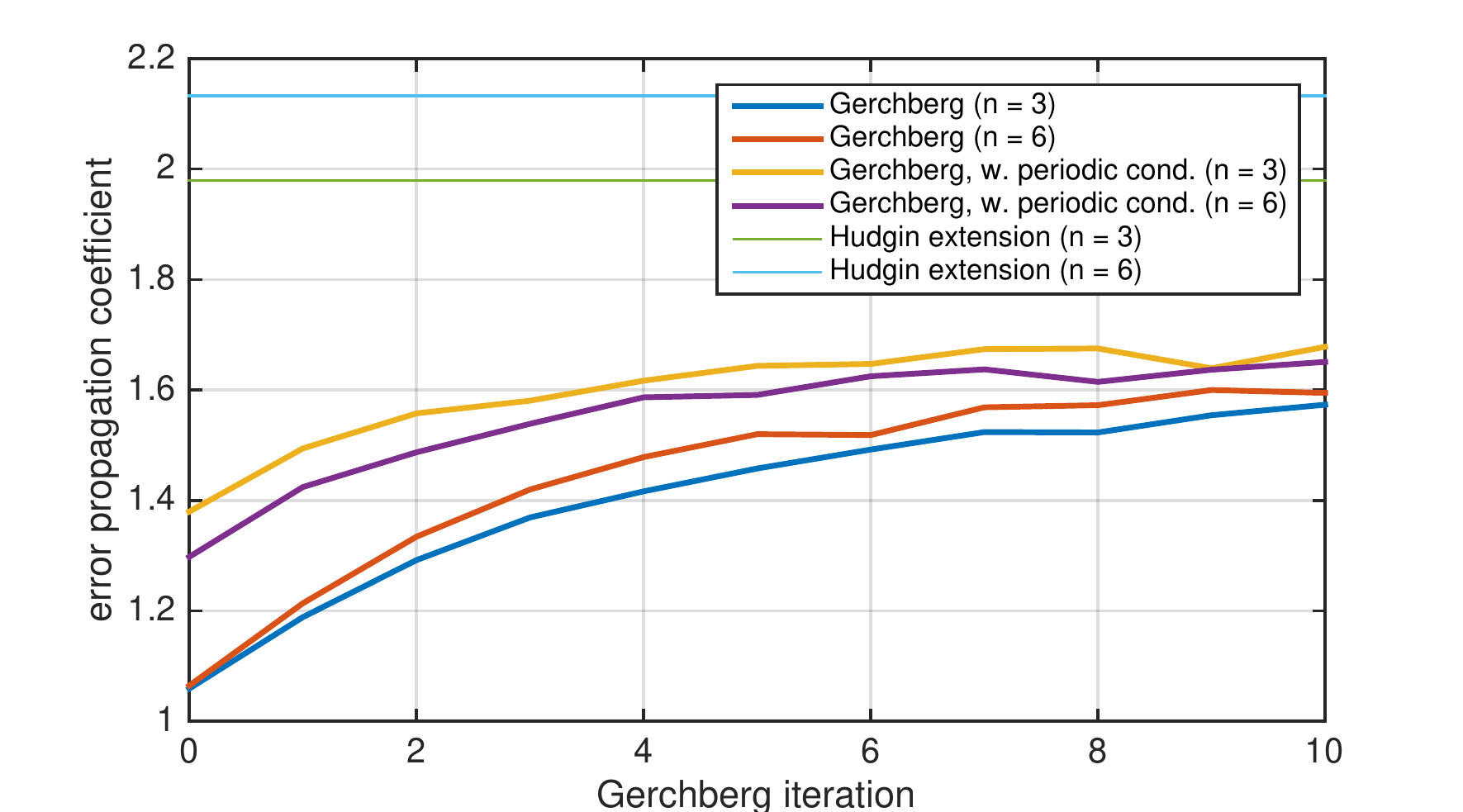}
    }
    \caption{Error propagation coefficient versus Gerchberg iteration
    		for different extension methods and extension windows ($n$).}
    \label{fig:OL_ErrProp}
\end{figure}

In addition to an improvement in Strehl the Gerchberg extension also
offers an improvement over the Hudgin extension in terms
of noise propagation, as illustrated in Fig.~\ref{fig:OL_ErrProp}.
Here the error propagation coefficient (as defined in ~\cite{Zou06}) versus Gerchberg iteration
is plotted, showing the improved performance of the Gerchberg.
The inclusion of the periodic condition does slightly increase
the propagation factor, as do the number of iterations.
Again we conclude in support of enforcing periodicity, as although
the noise propagation is higher for the same number of iterations 
we would require a large number of iterations without this condition to reach an acceptable Strehl,
resulting in a similar noise propagation but increased computation time.

\subsection{Closed loop}
\label{sec:CL}
\begin{figure}[t]
    \centerline{
      \includegraphics[scale=0.55]{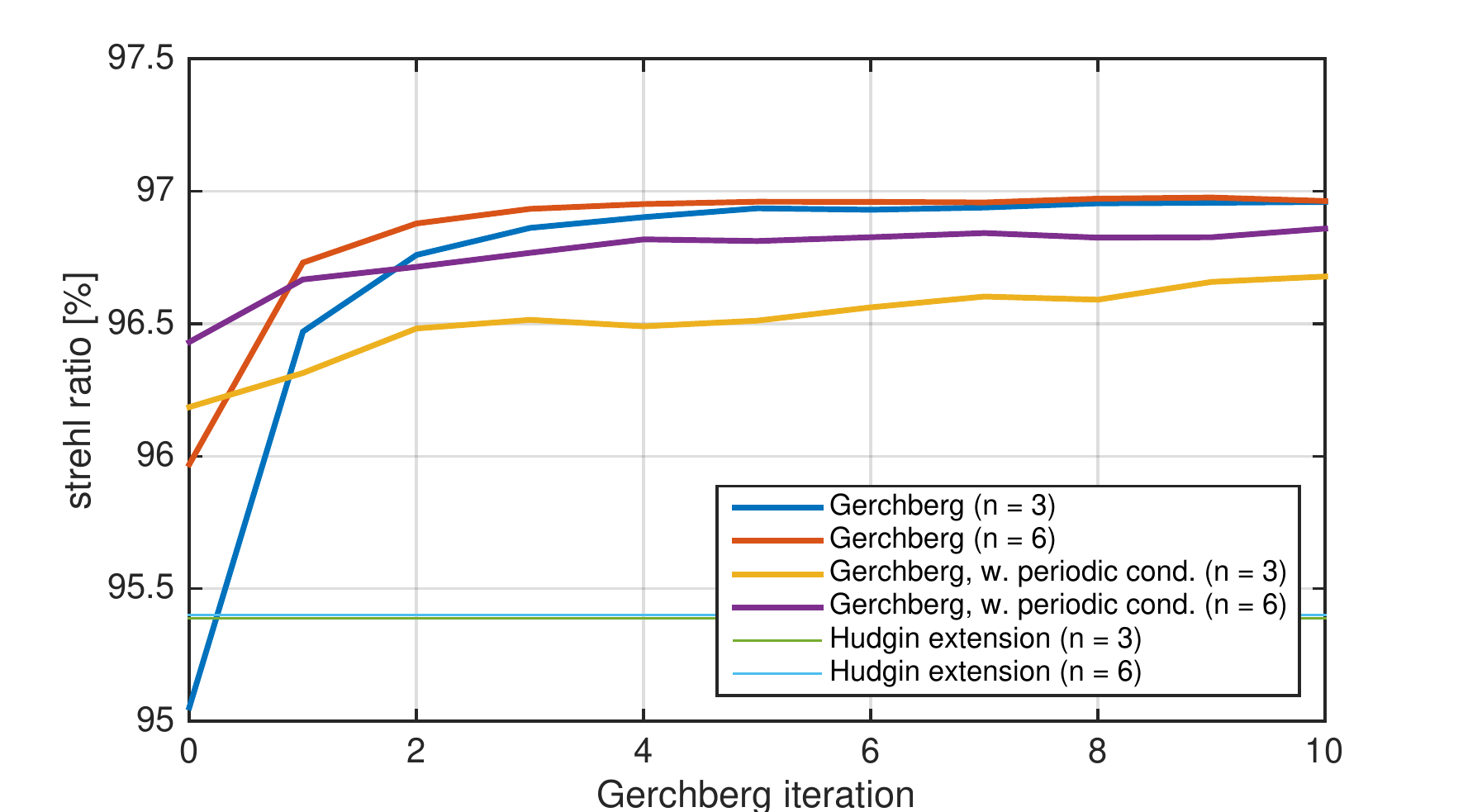}
    }
    \caption{Plots of Strehl ratio versus Gerchberg iteration for
    		closed loop simulations.  The results shown are
		for two different extension windows ($n$) and 3 different
		extension methods: 1) the Gerchberg extension; 2)
		the Gerchberg extension with periodicity enforced;
		and 3) the Hudgin extension. }
    \label{fig:CL_SRvsIter}
\end{figure}

In closed loop the residual phase is estimated, allowing for smaller
errors in the Fourier reconstruction due to the small nature of the aberrations.
In the closed loop simulations presented here the Strehl
is measured once the AO loop is closed, using a camera exposure of 200 closed 
loop time steps.
Simulating a closed loop case results in a Strehl of 81.6\%
when no extension method (or extension window) is applied.  The results in,
terms of Strehl, for different extension methods are summarised in Fig.~\ref{fig:CL_SRvsIter}.

In closed loop the Gerchberg delivers a greater Strehl than the
Hudgin extension and tends towards a high Strehl of $\sim 97\%$, close to
the estimated Strehl.
Temporal errors are included and
the estimation involves knowledge of the closed loop transfer function 
and time delay of the system~\cite{Hardy98} (1 frame delay in this case).  
Unlike in open loop
the number of iterations required for a stable Strehl both with and without the periodic condition
is small (1 or 2).  This is due to the nature of the closed loop, which
itself is iterative in nature, applying the
filtering process to the residual slopes.  In this way the larger errors seen in open
loop disappear.  

In closed loop the application of the periodic condition
in the Gerchberg actually causes a slight drop in performance.
The periodic constraint identified in Sec.~\ref{sec:gerchMeth}
is true for a no noise system of continuous (or concurrent) gradient measurements.
The discrete measurement process of the Shack-Hartmann
leads to the gradient of the average phase for each lenslet
which, due to the discrete measurement of the WFS spots at the focal plane
of each lenslet, is effectively taken across $d-d_{px}$, not the full
sub-aperture (see Sec.~\ref{Sec:FFTR}).
In this case the periodic condition is not the exact solution for a periodic phase.
In closed loop the Gerchberg extension (without the periodic condition) requires only a few iterations to
produce a consistent set of slope data, which due to the nature of the FFT
are periodic.  Enforcing the periodic condition is redundant and will add
a high frequency element to the data not representative of the measurement,
shifting the data away from its optimal extension
and introducing a slight error into the reconstruction.  Forgoing the periodic condition
achieves a greater Strehl after 1 iteration, allowing for maximising the
Strehl whilst avoiding the high frequency errors observed in open loop (see Sec.~\ref{sec:OL}).
Therefore for closed loop operation the periodic condition is not required.

As well as the increase in performance from the slope extensions
we again observe an increase in performance (in terms of Strehl) purely by increasing the size
of the reconstruction window (i.e. padding the slope data with zeros).  
This is illustrated in Fig.~\ref{fig:CL_SRvsIter} by the results for the Gerchberg method
with 0 iterations.  In theses cases the slopes have not been extended but the 
grid size has increased (by $n=3$ and $n=6$ points on 
both sides of the original data).
An increase in Strehl compared with the non-extended case (81.6\%)
is observed.  As shown in Fig.~\ref{fig:OL_PSDs} the errors induced by the
aperture are predominantly low spatial frequency errors.  Increasing
the size of the window shifts these errors into the low spatial frequencies defined
by the new window and out of those defined by the measurement data.
Further extension of the window (with and without enforced periodicity)
tends to a maximum Strehl of $\sim 96.5\%$ but does not reach the optimal performance
offered by the Gerchberg extension.  
In addition the results with zero padding
exhibit greater noise propagation and the development of unstable
waffle-like modes (as discussed in greater detail below).

\begin{figure}[h]
    \centerline{
      \includegraphics[scale=0.55]{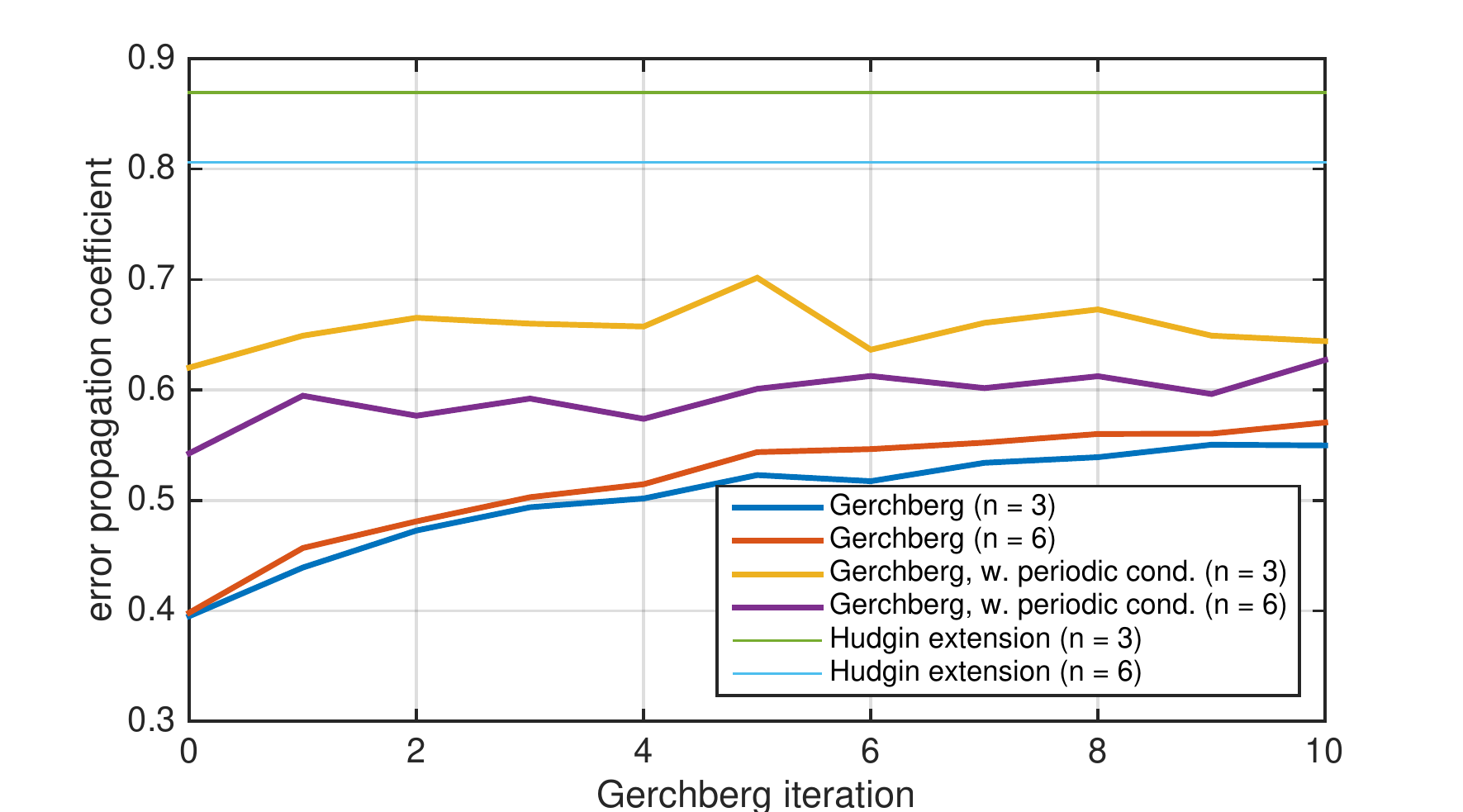}
    }
    \caption{Error propagation coefficient versus Gerchberg iteration
    		for different extension methods and extension windows ($n$)
		in end-to-end closed loop simulations.}
    \label{fig:CL_ErrProp}
\end{figure}

In addition to the improvements in Strehl offered by the Gerchberg the closed loop 
performance also demonstrates a reduced error
propagation compared to other extension methods.  These results are shown in Fig.~\ref{fig:CL_ErrProp}.
Again, as with the open loop results, the error propagation increases
with iteration and with the addition of the periodic condition.
For the closed loop however, we do not require the periodic condition
to maximise the Strehl with few iterations and so the Gerchberg
extension is optimal both in terms of Strehl and error propagation.

\begin{figure}[b]
    \centerline{
      \includegraphics[scale=0.55]{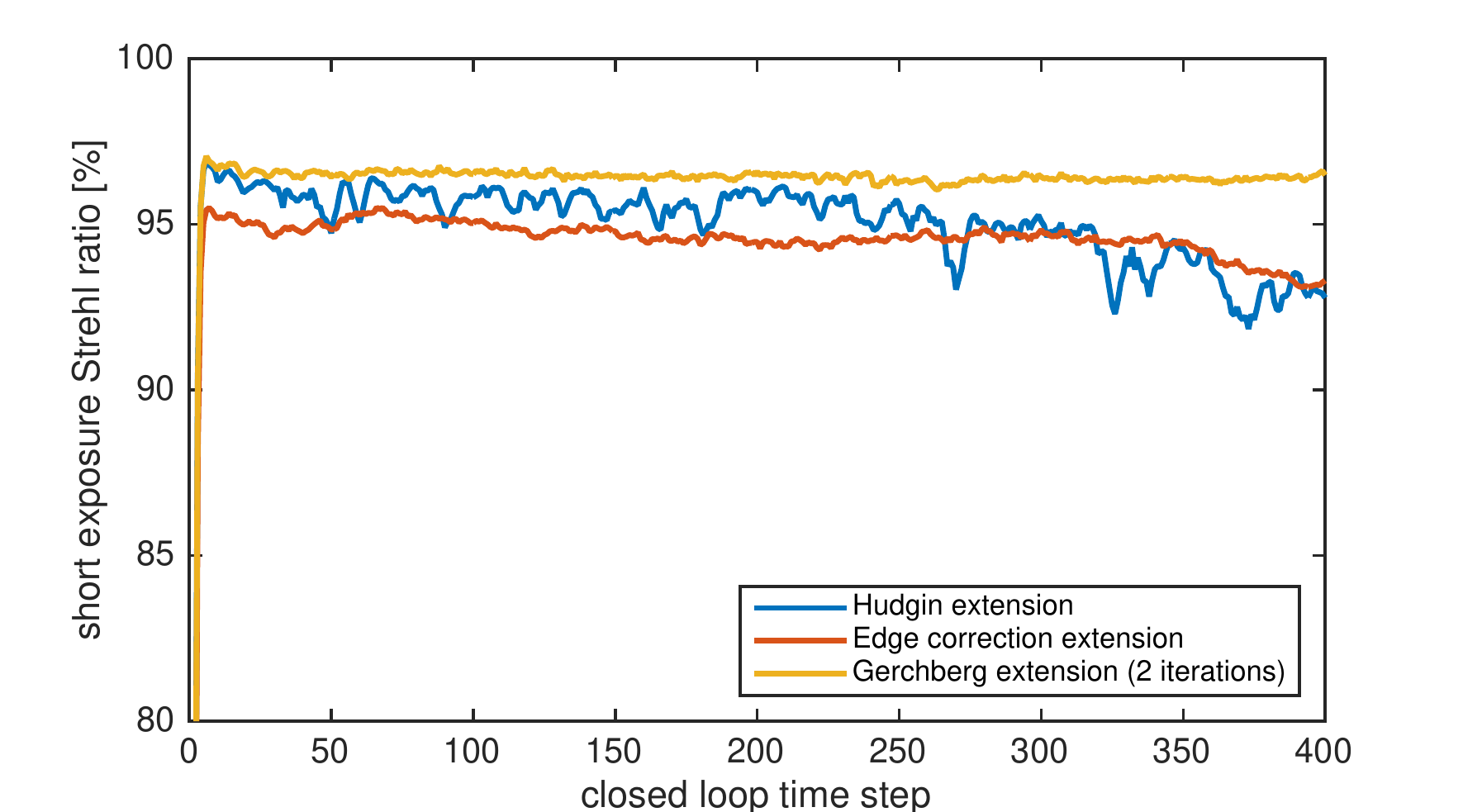}
    }
    \caption{Short exposure Strehl ratio vs. closed loop time step for 3 different extension methods:
    		1) Hudgin extension; 2) edge correction extension~\cite{Poyneer05}; and
		3) Gerchberg extension (2 iterations).  In all cases the local and global
		waffle modes have been filtered out.}
    \label{fig:CL_SRvsCLIter}
\end{figure}

The iterative process of the closed loop can result in small errors in the reconstruction
process propagating through the loop and producing large unstable modes.
In a standard Shack-Hartmann system these unstable modes are commonly the waffle modes.
In the closed loop results presented here we consider exposures
of 200 time steps.  When using the Hudgin extension
for longer exposures the beginnings of unstable behaviour were observed, even with
the application of global and local waffle filters.
This is illustrated in Fig.~\ref{fig:CL_SRvsCLIter}
where the short exposure Strehl for each time step of the closed loop is plotted
for different extension methods.  In this case we also include a comparison with another
extension method, the edge correction method~\cite{Poyneer05}, which only extends the
slopes into the data points immediately adjacent to the aperture.    In comparison with these two
methods the Gerchberg demonstrates very stable behaviour, even without the application of the 
waffle filters.  Although the results presented here use the Rigaut filter for phase reconstruction  
additional tests carried out using the \emph{Hudgin filter}~\cite{Correia14} with the Hudgin extension 
in closed loop demonstrated similar results and unstable behaviour.  
This issue of instability for the Hudgin and edge extension methods could be further
addressed by optimising the gain for each individual mode~\cite{Poyneer05}.
However, the optimisation of these extensions is not the subject of this paper.

The edge method
was not presented as the comparison method in this paper due to the better overall 
performance (in terms of Strehl) of the Hudgin extension (as shown in 
Fig.~\ref{fig:CL_SRvsCLIter}).  However, the edge extension demonstrates a less noisy
behaviour, compared to the Hudgin extension, whilst still being effected by the development
of unstable modes over time.  We can conclude that although there are other methods with the 
potential to increase the Strehl (increasing the extension window) or provide a more
stable/less noisy closed loop performance (edge extension) the Gerchberg extension
is the only method which optimises both and provides a very stable performance.

\begin{figure}[h]
    \centerline{
      \includegraphics[scale=0.5]{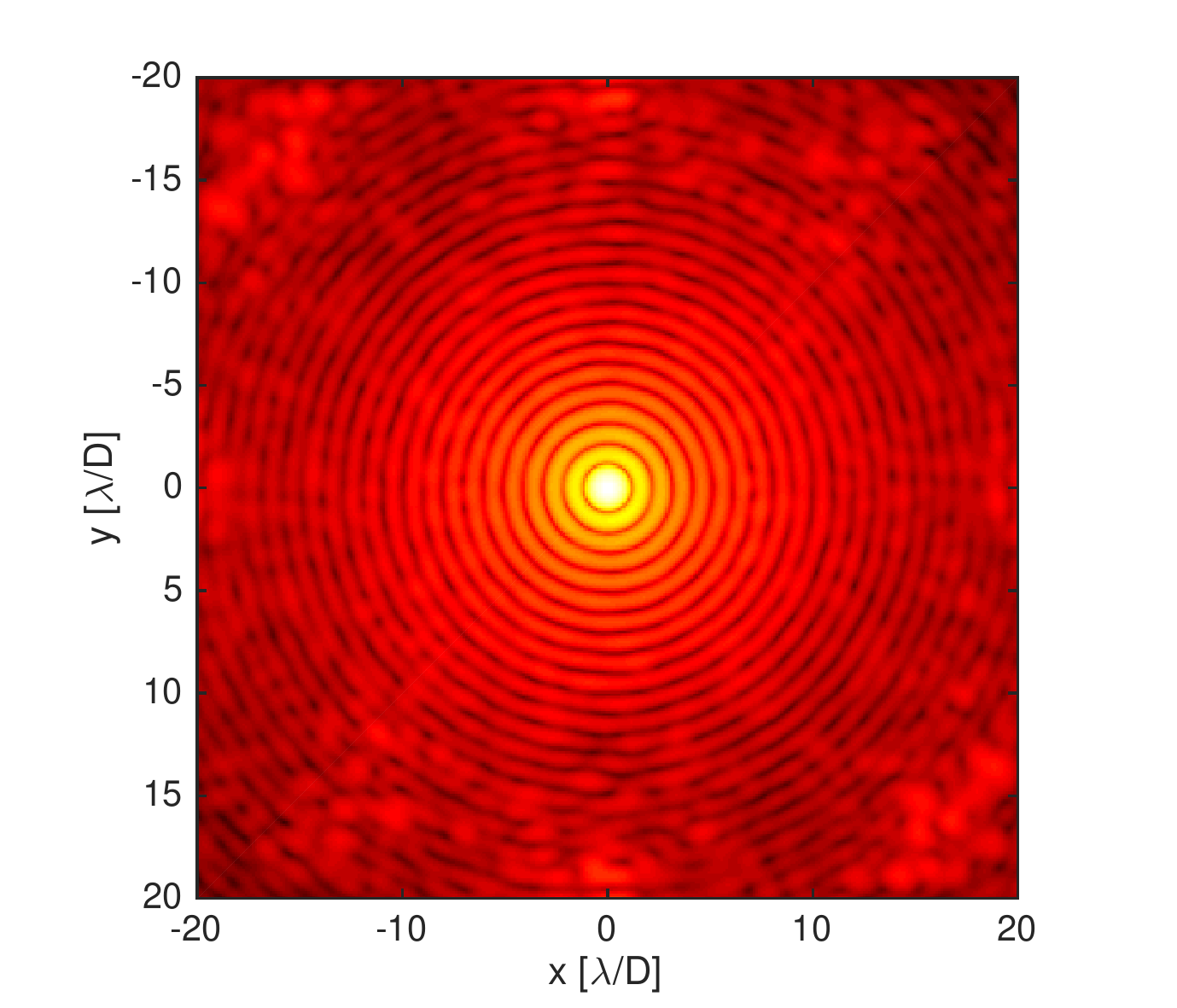}
      \includegraphics[scale=0.5]{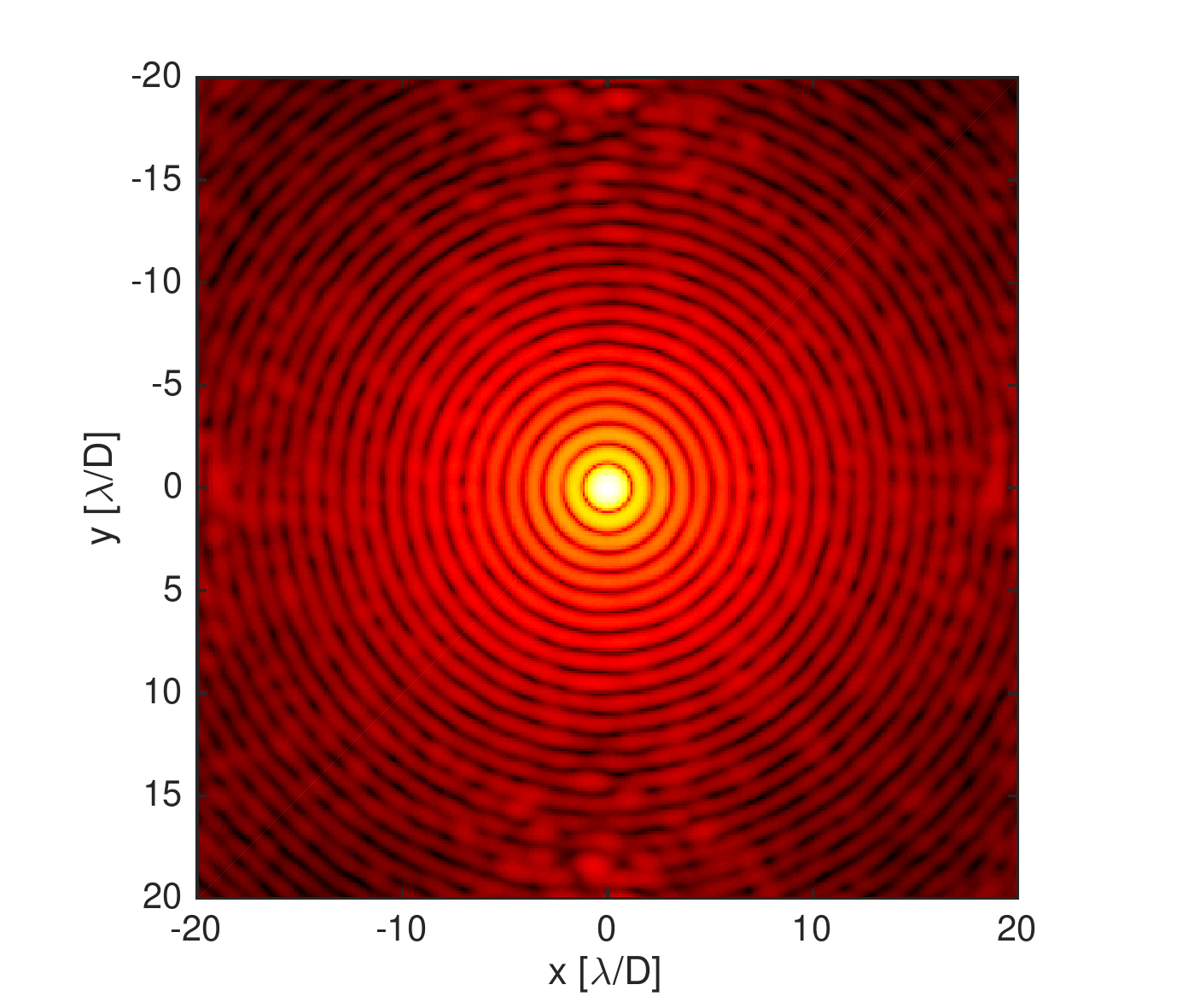}
    }
    \caption{Closed loop PSFs for Fourier reconstruction using different extension
    		methods.  {\bf Left:} Hudgin extension.  {\bf Right:} Gerchberg extension
		with $n_{\mathrm{iter.}}=2$.  Only the corrected region is shown and the same
		colour scheme is used for each image.}
    \label{fig:PSFs}
\end{figure}

Finally we consider the PSFs simulated for the different
extension methods in order to further compare the performance.  
Figure~\ref{fig:PSFs} shows the PSFs
generated over an exposure of 200 closed loop time steps
for the Hudgin extension (left) and 2 iterations
of the Gerchberg extension (right).  Only the corrected region is
shown.  The centre of the PSF appears similar for both cases.
However, the Gerchberg extension demonstrates an improved contrast towards the
edge of the correctable band, in the $10$ -- $20$\,$\frac{\lambda}{D}$
region and particularly in the corners.  This corresponds to
errors at high spatial frequency, particularly
waffle-like modes.

\section{Conclusions}

We propose a method for extending wave-front sensor data
outside of the aperture imposed by the telescope pupil based on an iterative Gerchberg routine.
Expanding on the work of Roddier\&Roddier~\cite{Roddier91} this extension method
involves recursive Fourier reconstruction of the phase across a rectangular
domain, extrapolating the slope measurements outside the bounds of the aperture.
The development of this extension method was motivated by a desire to achieve
improvements in Fourier wave-front reconstruction using anti-aliasing filters derived
in~\cite{Correia14}.  The Gerchberg method avoids some of the spatial
frequency errors and instabilities observed in other methods, allowing the full potential of such filters
to be realised.

Demonstration of this technique using end-to-end simulations 
has shown improved performance over
other extension methods.  In open loop an increase of $\sim3$\% Strehl (in K-band)
over previous methods is achieved, whilst noise propagation is reduced.
Such improvement can be achieved with minimal iterations of
the Gerchberg (1-2) when a periodic condition is enforced.
In closed loop gains in Strehl of $\sim 1.5$\% and a similar reduction in noise propagation
are observed, when compared with previous methods.
In the case of closed loop operation the periodic condition is not required to optimise
performance with 1-2 iterations and the Gerchberg method also exhibits 
extremely stable behaviour over time.
This corresponds to significant gains in rms ($\sim 40$\,nm) and an increase
in contrast for the final PSF ($10$ -- $20$\,$\frac{\lambda}{D}$).
This shows great potential for the development of such methods for future
high-contrast imagers, such as SPHERE 2.0.

The disadvantage of the Gerchberg method is the iterative process, which
will increase the computation time.  However, we require a limited number of
iterations: in both open and closed loop a single iteration improves performance
on existing methods.
For a single reconstruction process of an $N$ mode system the number of operations will
be $n_{\mathrm{iter.}}(2N\log(N) + 3N) + 2N\log(N) +N$, where $n_{iter.}$ is the number of Gerchberg
iterations, compared with $N^2$
operations for standard direct space reconstructors.  With the small number of iterations
required the increase in speed provided by the FFT is preserved
and we can expect a reduction in the number of operations greater than 1 order of magnitude
for future large AO systems.

Using the Gerchberg method we also anticipate further advantages
due to the ability to replicate the properties of the WFS measurements outside
the aperture without the sharp features introduced by other extension methods.  
This avoids introducing additional features to the reconstructed
phase, leading to errors particularly at high spatial frequency.  Here we have presented
results as applied to apertured data, but this method is potentially applicable to
other obstructions.  For example segmented telescopes, such as future ELTs, will
produce spatially discontinuous WFS data.  We foresee the use of the Gerchberg method
to extend the slope data between the segments, thus allowing for reconstruction
using Fourier methods.
In addition the Gerchberg formalism should be directly
applicable to other types of wave-front sensor, such as the Pyramid, whose
signals are not pure gradient measurements and so do not conform to
previously derived constraints.
Further investigations
into this method may also prove useful for the application of Fourier reconstruction over
non-cartesian grids, such as the actuator geometry proposed for the European Extremely
Large Telescope system HARMONI~\cite{Niranjan14}.

\section*{Funding}

The research leading to these results received the support of the A*MIDEX project 
(no. ANR-11-IDEX-0001- 02) funded by the Investissements d'Avenir French 
Government program and managed by the French National Research Agency (ANR).
This research was co-funded by the People Programme (Marie Curie Actions) of the European Union's Seventh Framework Programme (FP7/2007-2013) under REA grant agreement n. PCOFUND-GA-2013-609102, through the PRESTIGE programme coordinated by Campus France.

\section*{Acknowledgments}

All simulations presented are done using the object-oriented MALTAB AO 
simulator (OOMAO)~\cite{Conan14} freely available from \url{https://github.com/cmcorreia/LAM-Public}.



\end{document}